\documentclass[twocolumn,showpacs,amsmath,amssymb]{revtex4}

\usepackage{graphicx}
\usepackage{dcolumn}
\usepackage{bm}

\begin{document}

\title{Correlation Gap in Armchair Carbon Nanotubes}

\author{T. A. Gloor}
\email{Thomas.Gloor@ipt.unil.ch}	
\author{F. Mila}%
\affiliation{%
Institut de Physique Th\'eorique, Universit\'e de Lausanne, 
CH-1015 Lausanne, Switzerland
}%

\date{\today}

\begin{abstract}
We revisit the problem of the correlation gap in $(n,n)$ armchair
carbon nanotubes, that would be metallic in the absence of
electron-electron correlations. We attack the problem in the context
of a Hubbard model with on-site repulsion $U$ only, and we show that
the scaling of the gap as $\exp(-nt/U)$ predicted by Balents and
Fisher (Phys. Rev. B 55, R11973 (1997)), can only be valid if $U$ is not
too large, even for very large values of $n$. Using Hartree--Fock
calculations and Renormalisation Group arguments we derive the
scaling of the gap as a function of $n$ for a given value of
$U$. Possible applications for the magnitude of the correlation gap in
armchair carbon nanotubes will be discussed.
\end{abstract}

\maketitle

\section{Introduction}
In the last few years exciting experiments have been done on single
wall carbon nanotubes (SWCNT). It has been shown that SWCNT can behave
as a true quantum wire showing coherent transport \cite{Tans97},
Luttinger--liquid behaviour was proposed \cite{Bockrath99} and there are
indications for intrinsic superconductivity \cite{Tang01}. The typical
dimensions of the CNT used in these experiments are a few $\mathrm{nm}$ for the
\begin{figure}
\includegraphics[width=8cm]{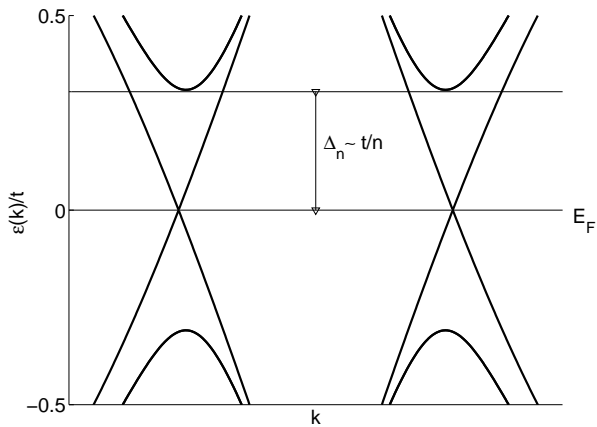}
\caption{\label{fig:fermiband}The energy bands near the Fermi level
for armchair CNT at half--filling.}
\end{figure}
diameter and several hundreds of $\mathrm{nm}$ for the length.
Treating CNT in a single particle picture, the electronic properties
depend strongly on chirality \cite{NT98}. In particular armchair CNT
are metalic and their Fermi surface at half--filling consists of
two points only. At this so-called Fermi points, bands with nearly
linear dispersion relations cross (Fig.~\ref{fig:fermiband}). This
property makes armchair CNT an ideal candidate for a one--dimensional
quantum wire forming a Luttinger liquid \cite{EggerGogolin98}.

In 1997 Balents and Fisher \cite{BalentsFisher97} considered the
problem of the correlation gap in half--filled $(n,n)$ armchair carbon
nanotubes. By excluding all but the lowest bands and using
an on--site Hubbard interaction $U$, they mapped the problem onto a
two--chain Hubbard model with an effective interaction $u_n=U/n$. For
this model it is known from renormalisation group (RG) calculations
(see \cite{BalentsFisher97} and references therein) that the
functional dependence of the gap $\Delta$ on $u_n$ and $t$ is given by
\begin{equation}
\label{eqn:BF}
\Delta \sim t\exp(-ct/u_n)\qquad\mbox{if } u_n\ll t.
\end{equation}
In other words, $(n,n)$ armchair CNT at half--filling with large
enough $n$ are predicted to be metallic for practical purposes since
the correlation gap is exponentially small.

In this paper we argue that the scaling law in Eqs.~(\ref{eqn:BF}) can
only be valid, even for large $n$, if at the same time the interaction
strength $U$ is not too large. This can already be seen from the $2D$
limit, i.e. when $n$ approaches infinity. In this limit we expect a
metal--insulator transition at some critical value $U_{cr}$ of the
interaction strength \cite{Sorella92, Furukawa01}. Even at finite $n$
the gap will 
be exponentially small only up to $U_{cr}$ but will grow linearly 
for larger values of $U$. In the following we derive the scaling law of the
gap as a function of $n$ and $U$. This gap is determined from
Hartree--Fock (H--F)
calculations where we take into account all the bands. We expect from
the H--F method that it produces the correct functional dependence in
the exponent of the correlation gap. This is the case for the Hubbard
model in one dimension where the Bethe Ansatz solution is reproduced
up to a prefactor \cite{MIT97}. On qualitative
grounds we show how we can get this scaling law from a RG argument. 

\section{The correlation gap from H--F calculations}
\begin{figure}
\includegraphics[width=8cm]{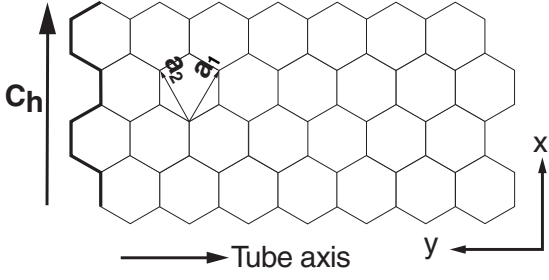}
\caption{\label{fig:honeycomb} A rectangular honeycomb lattice of
armchair type. The basis vectors are chosen to
be $\bm{a_1}=a/2(\sqrt{3},1) $ and
$\bm{a_2}=a/2(\sqrt{3},-1)$. $a=2.46\AA$ is the lattice constant of
graphene. The
chirality vector $\bm{C_h}$ points into the $x$ direction for armchair
CNT. The tube axis is perpendicular to it in the $y$ direction.}
\end{figure}

We are going to think about a SWCNT as a rectangular graphite
monolayer rolled up into a cylinder. This is equivalent to considering
a rectangular honeycomb lattice (Fig.~\ref{fig:honeycomb}) with the appropriate
periodic
boundary conditions. All the SWCNT can be classified by their chirality
vector $\bm{C_h} = n\bm{a_1} + m\bm{a_2}$, where $\bm{a_1}$ and $\bm{a_2}$
are the basis vector of the honeycomb lattice, while $n$ and $m$ are integers
with $m<n$ \cite{NT98}. $\bm{C_h}$ determines into which direction the
graphene layer is rolled up. As discussed in the introduction, the
armchair CNT are of particular interest since at half--filling they
are always metallic in the non-interacting electron approximation (see
Fig.~\ref{fig:fermiband}). Armchair CNT are characterized by chiral
vectors of the form  $\bm{C_h} = n\bm{a_1} + n\bm{a_2}$,
i.e. $n=m$. An example is shown in Fig.~\ref{fig:honeycomb}. In our
calculations we take the tubes to be 
very long and we use also periodic boundary conditions at the ends of
the tube. To determine the charge gap in $(n,n)$ armchair nanotubes,
we consider the following model with nearest neighbor hopping and
on--site Hubbard interaction at half--filling:
\begin{equation}
H = \sum_{\left\langle i,j\right\rangle \sigma}
\left(t_{ij}c^\dagger_{i\sigma} c_{j\sigma} +\mathrm{h.c.}\right) + U\sum_i n_{i\uparrow}n_{i\downarrow}
\end{equation}
$\sigma$ is the spin index and $i,j$ sum over the sites of a
rectangular armchair--type honeycomb lattice with periodic boundary
conditions. $c_{i\sigma}^\dagger$ are the fermion creation operators
and $n_{i\sigma} = c_{i\sigma}^\dagger c_{i\sigma}$. The hopping
integrals $t_{ij}$ are restricted to nearest neighbors and are taken
to be equal to a single $t$ for all the hoppings. In the case of
half--filling and on bipartite lattices with non--frustrated hopping,
the H--F Hamiltonian is \cite{MIT97}
\begin{eqnarray}
H &=& \sum_{\left\langle i,j\right\rangle \sigma}
t_{ij}c^\dagger_{i\sigma} c_{j\sigma}\nonumber + \mathrm{h.c.}\\&& + U\sum_i
\left[n_{i\uparrow}\langle n_{i\downarrow}\rangle
+\langle n_{i\uparrow}\rangle n_{i\downarrow} - 
\langle n_{i\uparrow}\rangle\langle n_{i\downarrow}\rangle\right]
\end{eqnarray}
with the expectation values given by
\begin{equation}
\left\langle n_{i\sigma}\right\rangle =
1/2\left(1+m(-1)^i\lambda_{\uparrow\sigma}\right)
\end{equation}
where $\lambda_{\sigma\sigma}=1$,  $\lambda_{\sigma -\sigma}=-1$ and
$m=\left|\left\langle n_{i\uparrow}\right\rangle-\left\langle
n_{i\downarrow}\right\rangle\right|$. This Hamiltonian can be
diagonalized by a Bogoliubov transformation and it yields the
self--consistent H--F equation for the sublattice magnetisation:
\begin{equation}
m=\frac{2}{N}\sum_{\bm{k}\in 1^{st}BZ} \frac{Um}{\sqrt{U^2m^2 +
4\epsilon^2(\bm{k})}} 
\end{equation}
$N$ is the total number of sites and $\epsilon(\bm{k})$ is the tight binding
dispersion relation for a single graphite layer given by:
\begin{eqnarray}
\epsilon^2(\bm{k})&=& 3t^2 + 2t^2\left[\cos(\bm{k}\cdot\bm{a_1}) +
\cos(\bm{k}\cdot\bm{a_2})\right.\nonumber\\ 
&&\left. +\cos(\bm{k}\cdot(\bm{a_1}-\bm{a_2}))\right]
\end{eqnarray}
Finally the gap is obtained from the sublattice magnetisation:
\begin{equation}
\Delta= \min_k \sqrt{U^2m^2 + 4\epsilon^2(k)} = Um
\end{equation}

\begin{figure}
\includegraphics[width=5.7cm]{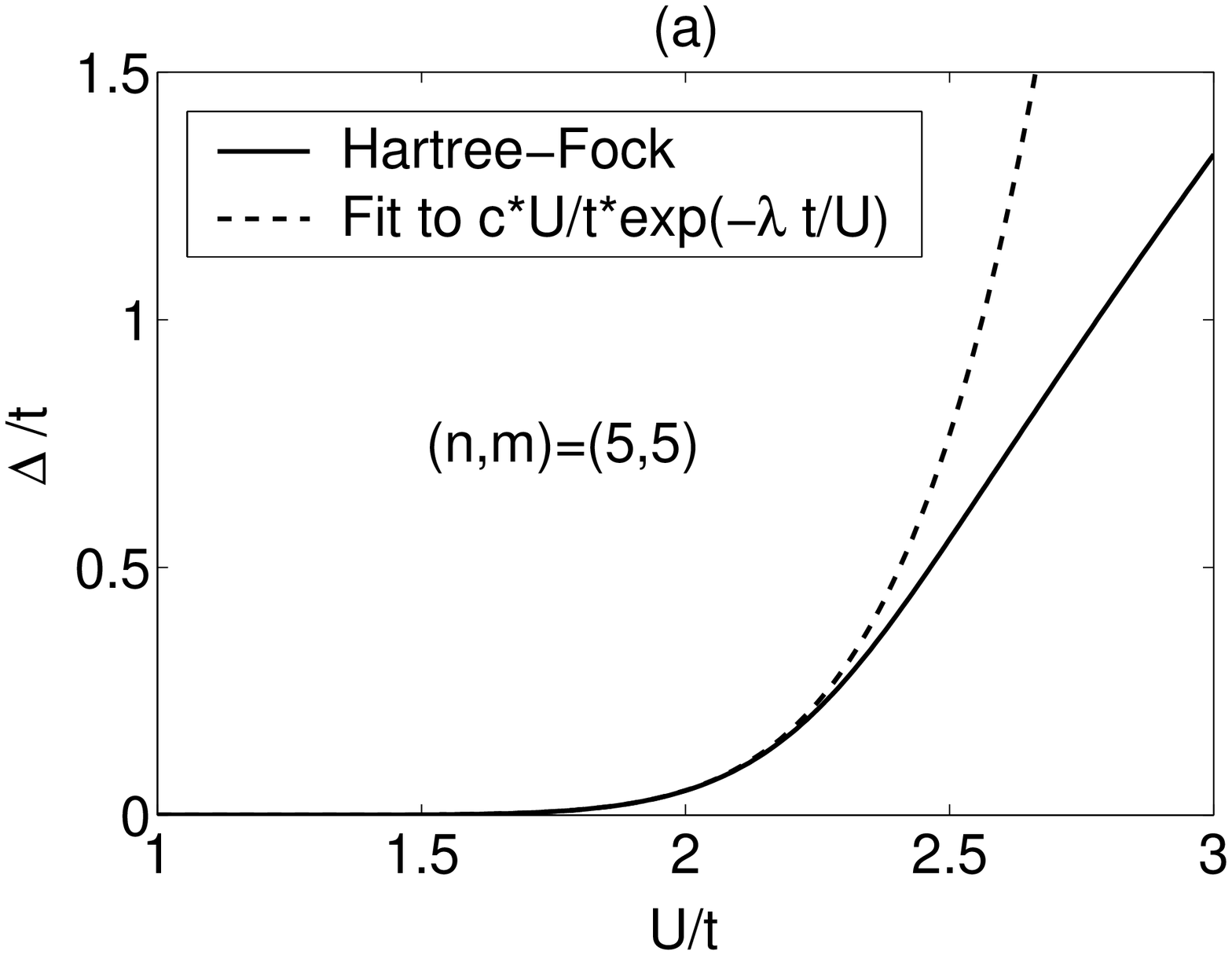}
\includegraphics[width=5.7cm]{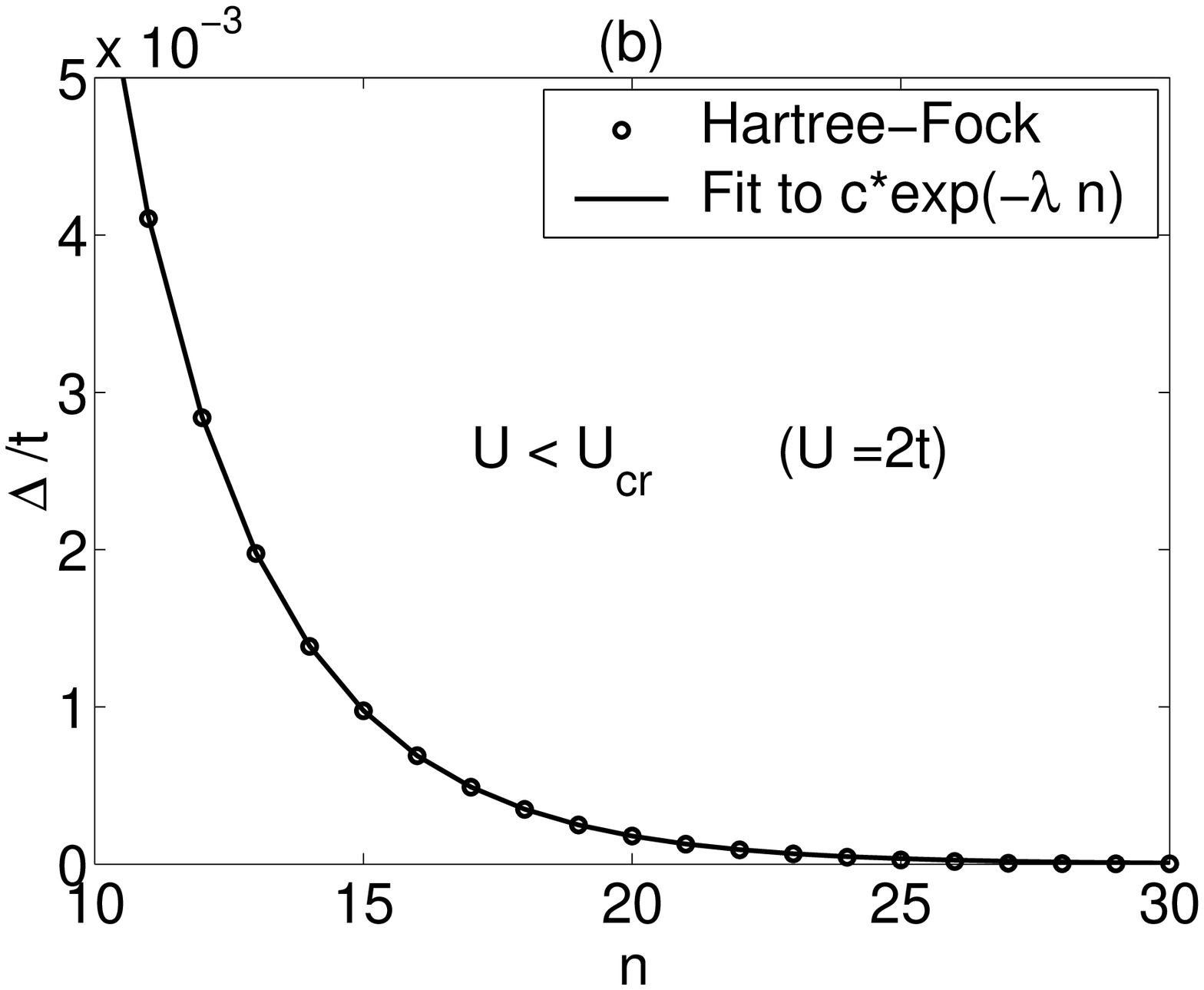}
\includegraphics[width=5.7cm]{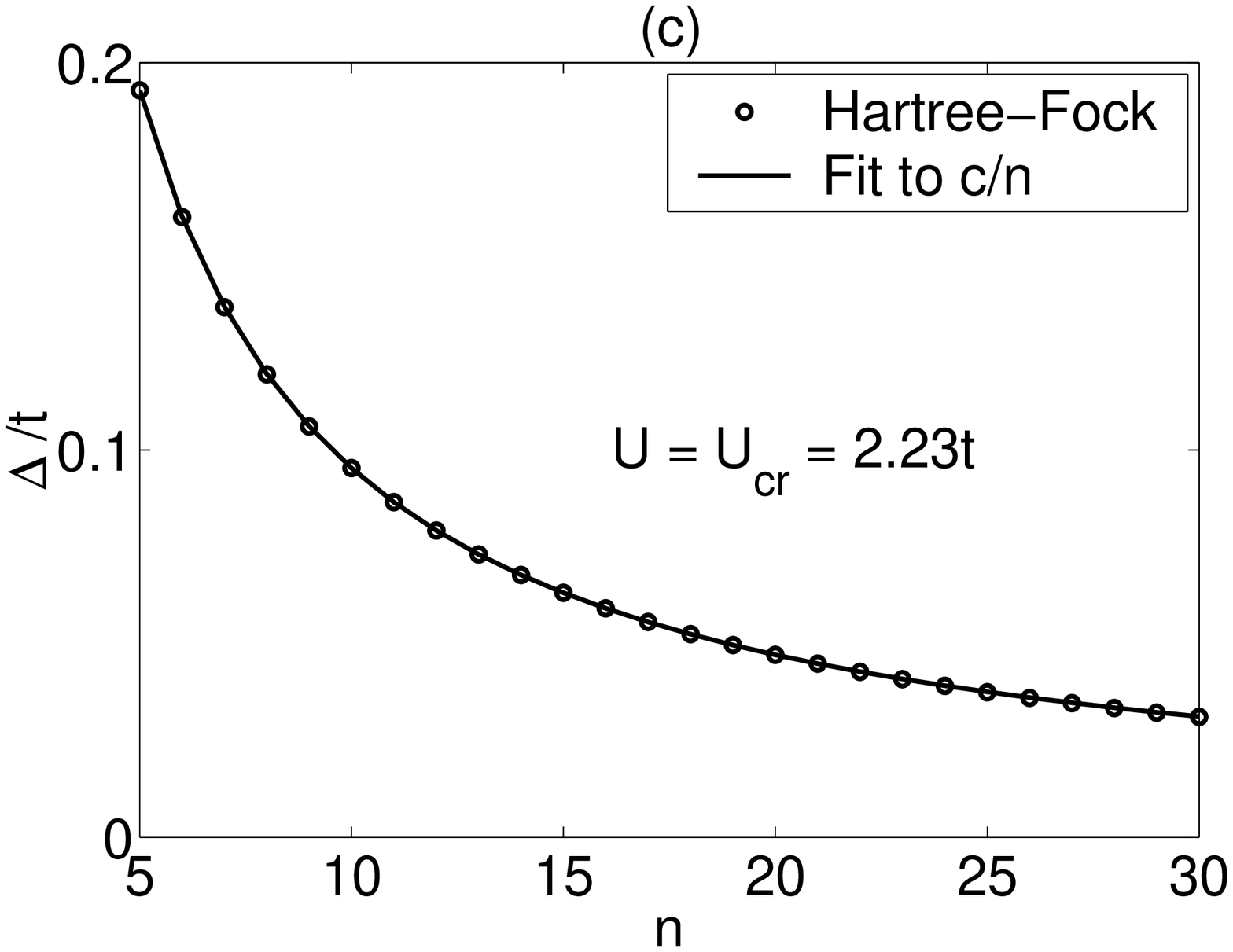}
\caption{\label{fig:HFfit}Numerical results for the correlation gap:
(a) The gap for a $(5,5)$ tube as a function of $U$. The solid line is
the H--F result. The dashed line is the fit to $c\,U/t\exp(-\lambda\,t/U)$
($c$, $\lambda$ are constants). (b) The gap as a function of the CNT size
$n$ at $U=2t$. The H--F results ($\circ$) are fitted to
$c\exp(-\lambda\,n)$ (solid line) (c) The same plot for
$U=U_{cr}^{HF}=2.23t$. The data fits the function $c/n$.}
\end{figure}
In Fig.~\ref{fig:HFfit} the numerical results for the correlation gap
are shown. We have two parameters to vary, the tube diameter which is
proportional to $n$ and the interaction strength $U$. We tried to fit
the H--F results to  $c\,U/t\exp(-\lambda\,t/U)$ when we varied $U$ and
to $c\exp(-\lambda\,n)$ for a variation of $n$. We see that these fits
break down at a critical value of the interaction strength
$U_{cr}^{HF}=2.23t$. There is no exponential decay beyond $U_{cr}^{HF}$. In
particular for $U=U_{cr}^{HF}$ we can fit the data to $c/n$, a power law. 
All our calculations can be summarized in the following scaling law
\begin{equation}
\Delta/t = 1/n\, \exp\left\{-\alpha n (t/U -
t/U_{cr}^{HF})\right\}
\end{equation}
with $\alpha = 5.44$. $U_{cr}^{HF}$ is identical to the critical value in
the H--F approximation to
open a gap in the two--dimensional honeycomb lattice. We see that if
$U$ is approaching $U_{cr}^{HF}$ the exponent is going to vanish and we are
left with a power law $\Delta=t/n$.

\section{RG--argument}
To compute the charge gap from a RG--calculation we need to compute
the RG--equation up to third order in the running coupling constant
\cite{LarkinSak77}. For armchair CNT the effective coupling
constant is given by $g=U/n$ \cite{BalentsFisher97} and we denote by
$\tilde g$ the renormalized coupling constant corresponding to the
momentum cutoff $\tilde D$. The third order equation is
\cite{PencMila94}
\begin{equation}
\frac{\mathrm{d}\tilde g}{\mathrm{d}\ln\tilde D} = -\frac{1}{\pi
v_c}\tilde g^2 + \frac{1}{2\pi^2v_c^2}\tilde g ^3
\end{equation}
with $v_c=v_F + U/(2\pi n)$. The RG flow has to be stopped at the
strong coupling region. This should occur when the momentum cutoff
$\tilde D$ corresponds to the energy scale of the charge gap
$\Delta_c$ \cite{LarkinSak77}. The flow of the running coupling
constant has to be stopped at the first relevant energy scale. In the
usual one--band picture this energy scale is just the kinetic energy
$\sim t$. But in the present problem there is another energy scale
entering before: It is the gap to the
next band which is of order $t/n$ . This argument leads to $\tilde
g(\Delta_c) \simeq t/n$. Now we can integrate out the degrees
of 
freedom far from the Fermi level, and we get
\begin{equation}
\int_{\Delta_c/v_c}^{\tilde D_0}\mathrm{d}(\ln\tilde D) = -
\int_{\tilde g (\Delta_c)}^g\mathrm{d}\tilde g\,
\left(\frac{\tilde g^2}{\pi
v_c} \left(1-\frac{\tilde g}{2\pi v_c}\right)\right)^{-1}
\end{equation}
where $\tilde D_0$ is the initial cutoff with $\tilde g (\tilde D_0) =
g = U/n$. To leading order in $g$ and $\tilde g (\Delta_c)$ we obtain
\begin{equation}
\ln\left(\frac{\Delta_c/v_c}{\tilde D_0}\right) \sim -\beta\frac{t}{g} +
\gamma\frac{t}{\tilde g(\Delta_c)}.
\end{equation}
Identifying the relevant energy scales we can write the charge gap as 
\begin{equation}
\Delta_c \sim \exp\left(-\beta t/(U/n) +\gamma n\right).
\end{equation}
We note that this argument leads to the H--F result.

\section{Discussion}
\begin{figure}
\vspace{0.5cm}
\includegraphics[width=7cm]{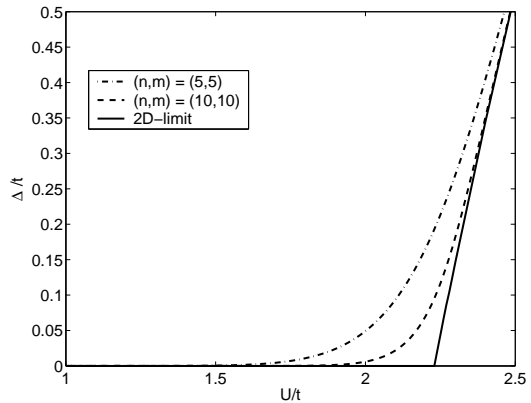}
\caption{\label{fig:gap}The correlation gap as a function of
interaction strength for $(5,5)$ CNT (dot-dashed line), $(10,10)$
(dashed line) and the 2D--limit (solid line).}
\end{figure}

Fig.~\ref{fig:gap} summarizes our results obtained from H--F
calculations and the RG argument on the correlation gap in armchair CNT. 
 We observed that there is an exponentially small gap only if
$U\ll U_{cr}^{HF}$, even for large $n$. $U_{cr}^{HF}$ is a critical value for the 
interaction strength. For values $U<U_{cr}^{HF}$ the correlation gap
follows an exponential scaling law. For $U>U_{cr}^{HF}$ the gap is growing
linearly. The actual value of the hopping integral $t$ in SWCNT is
about $2.4$ $\mathrm{eV}$ \cite{Mintmire92}. This value is large and as a
consequence we observe in the case $U\lesssim U_{cr}^{HF}$  a finite
gap which is of order $\mathrm{meV}$. As an example we look at a $(10,10)$ CNT
assuming an interaction strength of $U=2t$. In this case we obtain for
the gap a value of $0.006t$ which corresponds to an energy of $14$
$\mathrm{meV}$ and a temperature of $160$ $\mathrm{K}$. For smaller
CNT the gap is even larger.
\\
These results are based on two
approximations, namely the description of electron-electron interactions
in terms of a Hubbard model with only on-site repulsion, and the
calculation of the gap within Hartree-Fock theory. Let us now critically 
review these approximations.
\\
{\it Hubbard model:} It is by now well accepted that metallic nanotubes
behave as Luttinger liquids with an exponent controlled by the long-range
tail of the Coulomb repulsion and by the finite length of the tube \cite{EggerGogolin98}.
However, when it comes to calculating the correlation gap in half-filled
systems, the short-range part of the correlations plays a dominant role. 
This is most easily seen in the atomic limit where the hopping integrals 
are assumed to be vanishingly small. So let us assume for a moment that the
system is described by the Hamiltonian:
\begin{equation}
H=U\sum_i n_{i\uparrow}n_{i\downarrow} + \sum_{(i,j)} V_{ij} n_i n_j
\end{equation}
where $n_i=n_{i\uparrow}+n_{i\downarrow}$ is the total density, and
where $V_{ij}$ is the long-range part of the repulsion that does not
need to be specified. According to Mott's prescription 
to evaluate the gap \cite{MIT90}, 
one has to compare the energy of a uniform 
configuration with that of a configuration with a doubly occupied 
site and a hole far apart from each other. The only difference comes
from the energy of the electron that has been moved, and the energy
increase is precisely equal to $U$ since it still interacts with 
further neighbours in the same way. So, even in the presence of long-range
Coulomb repulsion, the value of the charge gap is controlled by the
on-site repulsion $U$ in the atomic limit. Of course, away from the atomic
limit, the long-range part of the Coulomb interaction will play a role.
To get a quantitative estimate of the charge gap in that case is a
very difficult problem though which has not been solved even in the simplest 
case of a pure one-dimensional model, but if anything the longer range
part of the Coulomb repulsion is expected to reinforce the tendency to 
localize the charge, hence to increase the charge gap. So to use a simple
Hubbard model with only on-site repulsion to calculate the correlation
gap of a half-filled system is a reasonable assumption, and the value is
probably an underestimate of the actual gap in the presence of the
long-range part of the Coulomb repulsion. For the present purpose
this is all we need since our main conclusion is to argue
that the gap might be larger than previously assumed. 
\\
{\it Hartree-Fock approximation:} In purely one-dimensional systems, Hartree-Fock
is known to reproduce correctly the exponential form of the gap as a 
function of $U$, but the prefactor of the exponential is wrong, and the gap
is overestimated by a factor $\sqrt{t/U}$. This is only a problem for
very small values of $U/t$, but in the range of interest to us, namely
not too far from $U_{cr}$, this is not a problem any more. In two dimensions,
very little is known precisely regarding the value of the correlation
gap for models without perfect nesting. Luckily enough, the Hubbard model
on the honeycomb lattice has been extensively studied by Monte Carlo 
simulations, and although the gap could not be calculated, the critical
value $U_{cr}$ has been determined by looking at the development of
magnetic long-range order \cite{Sorella92,Furukawa01}, and the value is around $U_{cr}=3.6 t$. This
is clearly larger than the Hartree-Fock result $U_{cr}^{\rm HF}=2.23 t$, but
the order of magnitude is the same. Since the gap is anyway expected to
grow linearly with $U$ beyond the critical value, the main result of 
Hartree-Fock that the gap in the nanotube geometry could take large
values if $U$ is not too far below $U_{cr}$ is expected to remain true
beyond Hartree-Fock. 
\\
The relevance of this calculation
depends crucially on the actual value of $U$ in nanotubes. What we need
here is the atomic value for carbon. The fact that graphite, a system
of very weakly coupled honeycomb planes, is a semi-metal and {\it not} 
an insulator just tells us that $U$ does not exceed $U_{cr}$. Besides, 
values often quoted for fullerenes are of no help since they concern 
the molecule $C_{60}$ and not the atomic value for carbon.
The most relevant value which we could find in the
literature was for polyacetylene where an on--site repulsion of
$5$--$10$ $\mathrm{eV}$ is generally used \cite{Jeckelmann95}. So we
believe that values in the range of $2t$ to $4t$ are to be
expected. In particular, a value of $U$ near the critical one seems
plausible.  Thus if in 
real CNT $U\lesssim U_{cr}$, a correlation 
gap of a few $\mathrm{meV}$ has to be present and a gap of such
magnitude could in principle be observed experimentally. 
\\
This of course will only be true for SWCNT at half--filling. Currently
available CNT seem to have some self--doping which shifts the
Fermi--level to an energy where several bands are cut. This turns the
CNT metallic independently of their chirality. There are ongoing
experimental efforts however identifying the nature of the doping and
producing undoped samples, and we are confident that it will be soon
be possible to test the conclusions of the present work.

\bibliography{nt}

\end{document}